%% file: main.tex
\documentclass{Interspeech}

\input{packages}

\interspeechcameraready 

\title{VietASR: Achieving Industry-level Vietnamese ASR with 50-hour labeled data and Large-Scale Speech Pretraining}

\author[affiliation={1,2}]{Jianheng}{Zhuo}
\author[affiliation={1}]{Yifan}{Yang}
\author[affiliation={2}]{Yiwen}{Shao}
\author[affiliation={2}]{Yong}{Xu}
\author[affiliation={2}]{Dong}{Yu}
\author[affiliation={1}]{Kai}{Yu}
\author[affiliation={1,\dagger}]{Xie}{Chen}

\affiliation{X-LANCE Lab}{School of Computer Science, MoE Key Lab of Artificial Intelligence, \\Shanghai Jiao Tong University, shanghai innovation institute}{China}
\affiliation{}{Tencent AI Lab}{USA}
\email{zzasdf@sjtu.edu.cn,yifanyeung@sjtu.edu.cn, yiwenyshao@global.tencent.com, yong.xu.ustc@gmail.com, DongYu@ieee.org, kai.yu@sjtu.edu.cn, chenxie95@sjtu.edu.cn}

\newcommand\blfootnote[1]{
  \begingroup
  \renewcommand\thefootnote{}\footnote{#1}
  \addtocounter{footnote}{-1}
  \endgroup
}

\begin{document}

\maketitle

\input{text/abstract}
\blfootnote{$\dagger$ Corresponding author. }
\input{text/introduction}

\input{text/related_work}
\input{text/method}
\input{text/experiment}

\input{text/limitation}

\input{text/conclusion}
\input{text/acknowledgement}

\bibliographystyle{IEEEtran}
\bibliography{mybib}

\end{document}

%% file: packages.tex
\usepackage{cite}

\usepackage{color,xcolor}
\usepackage{epsfig}
\usepackage{graphicx}
\usepackage{lipsum}
\usepackage{array}
\usepackage{booktabs}
\usepackage{colortbl}
\usepackage{multirow}
\usepackage{float}
\usepackage{caption}
\usepackage{adjustbox}
\usepackage{subfigure}
\usepackage{footnote}
\makesavenoteenv{tabular}
\makesavenoteenv{table}
\usepackage{makecell}
\usepackage{tablefootnote}

\usepackage{amsmath,amsfonts,amssymb,bm}
\usepackage[super]{nth}

\usepackage{changepage}
\usepackage{extramarks}
\usepackage{lastpage}
\usepackage{setspace}
\usepackage{soul}
\usepackage{xspace}

\usepackage{url}

\usepackage[ruled,vlined]{algorithm2e}
\usepackage{enumitem}
\usepackage{verbatim}
\usepackage{pifont}
\usepackage[acronyms]{glossaries}
\glsdisablehyper

%% file: text/abstract.tex
\begin{abstract}
Automatic speech recognition (ASR) has made remarkable progress but heavily relies on large-scale labeled data, which is scarce for low-resource languages like Vietnamese. While existing systems such as Whisper, USM, and MMS achieve promising performance, their efficacy remains inadequate in terms of training costs, latency, and accessibility. To address these issues, we propose VietASR, a novel ASR training pipeline that leverages vast amounts of unlabeled data and a small set of labeled data. Through multi-iteration ASR-biased self-supervised learning on a large-scale unlabeled dataset, VietASR offers a cost-effective and practical solution for enhancing ASR performance. Experiments demonstrate that pre-training on 70,000-hour unlabeled data and fine-tuning on merely 50-hour labeled data yield a lightweight but powerful ASR model. It outperforms Whisper Large-v3 and commercial ASR systems on real-world data. Our code and models will be open-sourced to facilitate research in low-resource ASR.
\end{abstract}

%% file: text/introduction.tex
\section{Introduction}
Advances in supervised learning~\cite{CTC, RNN_ASR, li2022recent} have led to significant improvements in the performance of automatic speech recognition (ASR). For mainstream languages like Chinese and English, it is feasible to collect large-scale labeled data for the model training, resulting in impressive ASR performance~\cite{whisper, canary, universial-1}. However, for low-resource languages like Vietnamese, acquiring sufficient labeled data remains a challenge. The availability of labeled data on the Internet differs by orders of magnitude compared to major languages like English. Therefore, developing ASR models that can effectively leverage limited labeled data for practical deployment is both critical and highly valuable for low-resource speech recognition.

To address this challenge, researchers have explored utilizing large-scale unlabeled speech data, which has become more and more accessible through publicly available sources. Existing methods for leveraging such data can be broadly categorized into two main approaches: self-supervised and semi-supervised learning. XLS-R~\cite{xls-r} is a family of models with up to 2 billion parameters, trained on nearly half a million hours of publicly available speech data across 128 languages. However, the training corpus of XLS-R is mainly focused on major languages within the Indo-European language family, providing insufficient support for many low-resource languages. Google USM~\cite{USM} trains ASR systems for over 100 languages by pre-training an encoder on web-scale unlabeled data, including 12 million hours of YouTube-based audio, and fine-tuning it on 90k hours of labeled data. It shows the potential of pre-training on a large scale of unlabeled data for multilingual ASR. However, Google USM produces a relatively large model (0.6$\sim$2B) and remains closed-source. MMS~\cite{MMS} increases the number of supported languages to over 1000, providing support for more languages with smaller speaker populations. However, the model size grows to 1 billion parameters, posing challenges for deployment and accessibility.
As for self-supervised models with low-resource language capabilities, they often have overwhelming numbers of parameters and use waveforms as the front-end, making them difficult to deploy as practical speech recognition models. GigaSpeech 2~\cite{yang2024gigaspeech2} leverages unlabeled data in a semi-supervised learning framework. It collects large-scale unlabeled audio for low-resource languages from YouTube, uses Whisper~\cite{whisper} to generate pseudo-labels, and refines the ASR model using noisy student training on the pseudo-labeled data. Given the large amounts of model parameters in Whisper, it is time-consuming to generate pseudo labels with Whisper, making it hard to scale to more data. In addition, the quality of the initial pseudo labels directly impacts the final ASR performance, where Whisper underperforms in many low-resource languages, making this approach less generalizable.

To overcome these limitations, we propose VietASR, an open-source ASR training pipeline tailored for low-resource languages\footnote{ Code is available at \url{https://github.com/zzasdf/VietASR}}.
VietASR leverages large-scale unlabeled data and limited labeled data to enhance model performance.
By optimizing the HuBERT pre-training for Zipformer~\cite{Zipformer} and introducing a supervised codebook to better align the pre-training with the ASR downstream task,
VietASR enables the training of a lightweight yet powerful ASR encoder with an Fbank front-end, improving both online and offline ASR models. 
Leveraging the efficient Zipformer architecture and the ScaledAdam optimizer, a single epoch of VietASR pre-training on 70,000 hours of audio takes only about 12 hours using 8 NVIDIA 32G V100 GPUs.
Our contributions can be summarized as follows:
\begin{itemize}[leftmargin=*,noitemsep]
    \item  We develop VietASR, an efficient large-scale speech pre-training pipeline for low-resource languages, suitable for both online and offline ASR models. By optimizing the HuBERT pre-training algorithm for models using Fbank front-end, VietASR can effectively leverage in-the-wild audio with automatic segmentation to train deployable ASR models.
    \item We propose a method to provide supervised signals for the pre-training process using only a small amount of labeled data. By leveraging a supervised codebook trained on limited labeled data, we significantly enhance the performance of pre-training on downstream ASR tasks.
    \item Experimental results show that with VietASR pre-training, a 68M Vietnamese ASR model fine-tuned with just 50 hours of labeled data, outperforms the Whisper-large-v3 (1.5B) and commercial models.
\end{itemize}


%% file: text/related_work.tex
\section{HuBERT and Its Variants}
HuBERT~\cite{Hubert}, short for Hidden-Unit BERT, is a self-supervised learning (SSL) method capable of leveraging masked prediction to learn robust speech representations.
It adopts the model architecture of wav2vec 2.0~\cite{wav2vec2}, comprising a CNN feature extractor to convert waveform into a local feature, and a transformer as the context network.
HuBERT takes raw waveforms as input, applies masking to the feature extractor’s output, and feeds the masked features into the transformer encoder to predict the hidden units of the masked regions.
HuBERT training involves multiple iterations. In the first iteration, SSL targets are generated by applying k-means clustering to MFCC features, while subsequent iterations improve label quality by re-clustering features from the pre-trained model.

Many works have been proposed to optimize HuBERT in terms of SSL targets~\cite{PBERT, polybert, HuBERT-AP, hubert_academic, ASRBERT, bias_asr, ctcbert} and training efficiency~\cite{MelHubert, FastHubert, yang2024k2ssl}.
To better align pre-training with the downstream ASR task, several studies~\cite{PBERT, hubert_academic, ASRBERT, bias_asr} have explored ASR-biased HuBERT, where the learning target is derived from a supervised model rather than being purely self-supervised.
PBERT~\cite{PBERT} introduces phoneme-level alignment with an additional phoneme recognizer to create a supervision-guided codebook. Another work~\cite{hubert_academic} leverages an ASR model trained on a large-scale labeled dataset as a feature extractor, applying k-means clustering to the extracted features for the SSL target, accelerating pretraining convergence. However, obtaining such a strong ASR model for low-resource languages is challenging.
Methods in \cite{ASRBERT, bias_asr} use feature clusters from a fine-tuned model as targets in the second iteration of HuBERT pre-training. Although these methods eliminate the need for a supervised ASR model, they still incorporate spectrum clustering in the first iteration, which is less aligned with the ASR task and computationally expensive.
To improve training efficiency, MelHuBERT~\cite{MelHubert} replaces raw waveforms with Fbank features and simplifies the HuBERT loss function. Fast-HuBERT~\cite{FastHubert} further enhances training efficiency by using Fbank with larger frame shifts.

%% file: text/method.tex
\section{VietASR}
To make the most of unlabeled data for building a speech recognition model suitable for real-world applications in low-resource languages, we propose the VietASR training pipeline. The design of VietASR focuses on two key aspects: 
\begin{itemize}[leftmargin=*,noitemsep]
    \item  Adapting HuBERT-style multi-iteration pre-training to  Zipformer encoder for better efficiency and practicality.
    \item Introducing a supervised codebook as the pre-training target to better align pre-training with downstream ASR tasks.
\vspace{-0.5em}
\end{itemize}

\subsection{HuBERT-style Pre-training with Zipformer}
Transformer~\cite{transformer} and Conformer~\cite{conformer} are widely used for speech pre-training. These pre-trained models are typically large in size and take raw waveforms as input\cite{MMS, xls-r, wav2vec2}, rather than the spectrograms commonly used in deployed speech recognition models. Zipformer~\cite{Zipformer} is a Transformer variant. Its high performance on speech recognition, lightweight architecture, and Fbank front-end make it a better backbone for practical speech recognition models. To this end, we adapted the HuBERT pre-training algorithm for Zipformer, enabling it to leverage large-scale unlabeled data for pre-training.

\subsubsection{Zipformer Backbone}
The Zipformer encoder consists of a convolution-based module, Conv-Embed, and a series of Zipformer blocks. Conv-Embed takes Fbank as input, downsamples it, and passes it to the Zipformer blocks. These blocks are organized into stacks, forming a U-Net structure where the middle stacks operate at lower frame rates. This U-Net-like architecture enables Zipformer to learn temporal representations at different resolutions more efficiently.

During pre-training, a linear projection layer is added after the encoder to predict labels for masked regions. For fine-tuning, a joiner network and a stateless predictor~\cite{stateless_predictor} are added after the Zipformer blocks to form an RNN-T architecture~\cite{transducer}. k2SSL~\cite{yang2024k2ssl} also uses Zipformer as the backbone for SSL, but it replaces the Fbank front-end with waveform, thus introducing a large CNN front-end, making the model less practical. 

\subsubsection{Mask Strategy}
The original HuBERT takes raw waveform as input \cite{Hubert}, and uses a CNN to aggregate local information from the waveform into high-dimensional features. The mask used in MLM pre-training is applied to the high-dimensional features. However, Zipformer uses Fbank as the model input, which is a multi-dimensional feature that already contains local information, so we apply masking to the Fbank features directly as in \cite{FastHubert}.



\subsubsection{Loss Function}
Since the original HuBERT prediction function requires a large amount of memory for updating the codeword embeddings and performing cosine similarity calculations, we replace it with a simplified function, following \cite{MelHubert} and \cite{FastHubert}, and adopt cross-entropy loss as the loss function,  The loss function is given by 
\begin{equation}
    f(\widetilde{X}, \mathbf{c}) = -\sum_{masked\ t}\log \frac{\exp(z_{tc_t})}{\sum_{i=1}^C \exp(z_{ti})}
\end{equation}
\begin{equation}
    \mathbf{o} = \bf{Encoder}(\widetilde{X}) = (o_1,o_2,\cdots, o_T)
\end{equation}
\begin{equation}
    \mathbf{z}_t = Ao_t/\tau = (z_{t1},z_{t2},\cdots, z_{tC})
\end{equation}
here $\widetilde{X}$ denotes the input with mask, $\mathbf{c} = (c_1,c_2,\cdots, c_T)$ denotes the cluster label,
$A$ denotes a projection matrix, and $\tau$ denotes a temperature parameter.

\begin{figure*}[t]
\centering
\includegraphics[width=\textwidth]{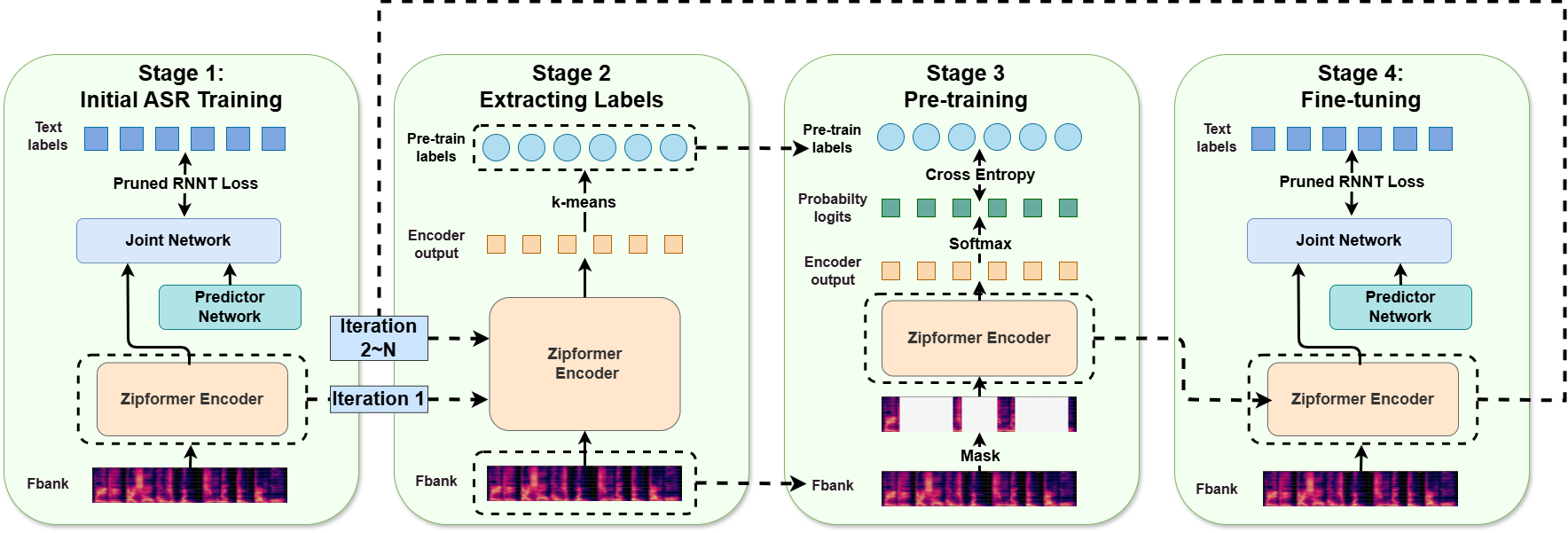}
\caption{An overview of the proposed VietASR. The training pipeline consists of four stages: (1) initial ASR training, (2) label extraction, (3) pre-training, and (4) fine-tuning. Components within the dashed box are reused in the next stage. In the first iteration, Stage 2 reuses the encoder trained in Stage 1 for label extraction; in subsequent iterations, Stage 2 reuses the encoder from Stage 4 of the previous iteration, enabling iterative refinement and continuous model improvement.}
\label{fig:pipeline}
\vspace{-1em}
\end{figure*}

\subsection{Speech Pre-training Pipeline with ASR-biased SSL}
The original HuBERT uses k-means clustering labels on unsupervised data only. The k-means labels are not directly related to phonemes or semantics, which makes it difficult for the model to learn ASR-relevant features during pre-training, leading to low training efficiency. To better align HuBERT's pre-training with ASR tasks and improve efficiency, a codebook derived from an ASR model is proposed to replace HuBERT’s original learning target.

While obtaining large-scale transcribed audio data for a low-resource language is challenging and costly, acquiring dozens of hours of transcribed data is often feasible. Training a high-quality ASR model solely on limited paired data is often inadequate; however, even a weak ASR model encodes knowledge on speech-text alignment, thus making the pre-training process more relevant to downstream ASR tasks. To leverage this insight, we use k-means clustering on the output of an ASR model trained with limited labeled data as the pre-training target for VietASR.

The full VietASR pipeline includes four stages.
In Stage 1, a Zipformer ASR model is trained on a small set of supervised data using the pruned RNN-T loss~\cite{Pruned-RNN-T}.
In Stage 2, k-means clustering is applied to the final-layer features of the trained Zipformer encoder to generate labels for speech pre-training.
Stage 3 involves speech pre-training using the ASR-biased labels from Stage 2, optimizing with a cross-entropy loss.
In Stage 4, the pre-trained encoder is fine-tuned using pruned RNNT loss.

Each subsequent iteration begins at Stage 2, replacing the ASR encoder from Stage 1 with the fine-tuned encoder from the previous iteration's Stage 4, and then follows the same pipeline as iteration 1. Our method is similar to \cite{hubert_academic, ASRBERT, bias_asr}, but eliminates the need for a strong ASR feature encoder and circumvents the inefficiencies of the first iteration, which relies on MFCC clustering labels.

%% file: text/experiment.tex
\section{Experiments}
\label{sec:experiment}
\subsection{Experimental Setups}
\noindent\textbf{Dataset}\quad
We collect approximately 70,000 hours of unlabeled audio from YouTube and apply  FunASR VAD~\cite{gao2023funasr} for automatic segmentation. 50-hour audio from YouTube is adopted for fine-tuning, which is manually transcribed by professionals. 
Model evaluation is conducted on three public Vietnamese test sets, consisting of
GigaSpeech~2\cite{yang2024gigaspeech2}, Common Voice 17.0~\cite{ardila-etal-2020-common}, and FLEURS~\cite{conneau2023fleurs}. A summary of the dataset is presented in Table \ref{tab:dataset}.

\begin{table}[ht]
\centering
\caption{Overview of the dataset used in the experiment.}
\label{dataset}
\renewcommand{\arraystretch}{1.1}
\resizebox{\linewidth}{!}{
\renewcommand\tabcolsep{12.0pt}
\begin{tabular}{cccc}
\toprule
\textbf{Category} & \textbf{Dataset} & \textbf{Duration (h)} & \textbf{Source} \\
\midrule
Pre-training Set & YouTube Unlabeled Data & 73000 & YouTube \\
\hline
Fine-tuning Set & Manual Transcribed Data & 50 &  YouTube \\
\hline
\multirow{3}{*}{Test Set} & GigaSpeech 2 TEST & 11 & YouTube \\
& Common Voice TEST & 1 & Read-Aloud Data  \\
& FLEURS TEST & 3 & Read-Aloud Data  \\
\bottomrule
\end{tabular}}
\label{tab:dataset}
\vspace{-1em}
\end{table}

\noindent\textbf{Extracting Labels}\quad
We evaluate our training pipeline on a multi-iteration setting and use a cluster size of 500 for the k-means clustering. For the first iteration, we train an ASR model on only 50h hours of labeled data and run k-means clustering over the encoder output to extract pre-training labels. For the rest iterations, we finetune the last checkpoint from the previous iteration on the 50h dataset and apply k-means clustering on the encoder output.

\noindent\textbf{Pre-training}\quad
We use the encoder of the Zipformer base as the encoder backbone, which consists of 6 stacks of Zipformer blocks, and a downsampling factor of $(1,2,4,8,4,2)$ for each stack.
We apply the mask on the time dimension with a configure like \cite{Hubert}, the mask span set to 10, and 8\% of the Fbank frames are randomly selected as mask start. For optimization, we use ScaledAdam with $\beta = (0.9, 0.98)$ as the optimizer and Eden as the scheduler.
All of the experiments of pre-training are conducted on 8 NVIDIA V100 GPUs with 1000s audio in a single batch. Due to the lower label quality in the first iteration, extending its training time yields limited benefits. Thus, we train the first iteration for 9 epochs, while each subsequent iteration undergoes 18 epochs. More details can be found in Table~\ref{tab:training_detail}

\definecolor{darkgreen}{rgb}{0,0.5,0}

\begin{table}[th]
\vspace{-1em}
\centering
\caption{Training details for each iteration.}

\renewcommand{\arraystretch}{1.1}
\renewcommand\tabcolsep{6pt}
\resizebox{\linewidth}{!}{
\begin{tabular}{lcccc}
\toprule
 & \textbf{Epochs} & \textbf{Steps} & \textbf{Device} & \textbf{\makecell{GPU \\ Hours}} \\
\midrule
Iteration 1 & 9  & 286k & 8 NVIDIA 32GB V100 & 8 $\times$ 96h  \\ 
Iteration 2 & 18 & 573k & 8 NVIDIA 32GB V100 & 8 $\times$ 194h \\ 
Iteration 3 & 18 & 573k & 8 NVIDIA 32GB V100 & 8 $\times$ 194h \\ 
\hline
\multicolumn{4}{l}{Overall Elapsed Time} &  \multicolumn{1}{c}{$\sim$20 days} \\
\bottomrule
\end{tabular}}
\label{tab:training_detail}
\vspace{-1em}
\end{table}

\noindent\textbf{Fine-tuning}\quad
We leverage pruned RNN-T loss~\cite{Pruned-RNN-T} to fine-tune the pre-trained encoder. A joint network and a predictor network are added to the encoder for fine-tuning. The joint network has a hidden size of 512. The predictor network is a stateless decoder with a hidden size of 512 and a context size of 2. 500-class Byte Pair Encoding (BPE)~\cite{BPE} word pieces are selected as the classification units.
The fine-tuning is conducted on 4 NVIDIA V100 GPUs, with 1000s audio in a single batch.
If not specifically mentioned, we use constrained beam search~\cite{modified_beam_search} with a beam size of 4 for ASR decoding.

\subsection{Comparison to Open-Source and Commercial Systems}

We compare the performance of VietASR with some open-sourced system and commercial systems, including Whisper base, Whisper large-v3~\cite{whisper}, Google USM~\cite{USM}, MMS L1107~\cite{MMS}, Azure Speech CLI, and the 68 M and 152 M model from GigaSpeech 2.0~\cite{yang2024gigaspeech2}. We adopted Azure Speech CLI of version 1.37, and Chirp Speech-to-Text v2 model For Google USM.

The results are presented in Table \ref{tab:main}. Compared to the non-pretrained Zipformer, VietASR achieves significant improvements, with WER consistently decreasing as the number of iterations increases. This demonstrates the effectiveness of the VietASR pipeline in learning labels suitable for downstream ASR tasks through iterative training. The iteration 3 model of VietASR achieves the lowest average WER, 29.5\% lower than the second-best model.

On the GigaSpeech 2 test set, VietASR surpasses all the other models, including the 1542M-parameter Whisper Large v3 and commercial systems, significantly better than on the other two test sets. We attribute this to the domain similarity between the pre-training dataset and the GigaSpeech 2 test set, as both are collected from YouTube. Although the domains are not entirely consistent, on the Common Voice and FLEURS test set, VietASR still achieves the second-best performance, comparable to commercial systems, demonstrating its robustness.


\begin{table}[ht]
\centering
\caption{Comparison of VietASR with mainstream open-source and commercial models. The average WER (Avg) is weighted by the word count of each test set. The lowest WER is highlighted in \textbf{bold}, and the second-lowest WER result is \underline{underlined}. For VietASR, only the third iteration is considered for ranking.}

\renewcommand{\arraystretch}{1.1}
\resizebox{\linewidth}{!}{
\renewcommand\tabcolsep{2.0pt}
\begin{tabular}{lcccc>{\columncolor[HTML]{F4F4F4}}c}
\toprule
\textbf{System} & \textbf{\makecell{\# Params \\ (M)}} & \textbf{\makecell{Giga- \\ Speech 2}} & \textbf{\makecell{Common \\ Voice}} & \textbf{FLEURS} & \textbf{Avg} \\
\midrule
\textbf{Open Source System} \\
Whisper large-v3 & 1542 & 17.94 & 13.74 & \textbf{8.59} & 16.44 \\
Whisper base & 72 & 39.88 & 44.07 & 40.41 & 40.16 \\
MMS L1107 & 964 & 46.62 & 43.88 & 55.35 & 47.67 \\
GigaSpeech 2 & 68 & 14.72 & 18.81 & 13.50 & 14.75 \\
GigaSpeech 2 & 152 & 12.83 & 14.43 & 11.59 & 12.74 \\
\hline
\textbf{Commercial System} \\
Google USM & - & 13.28 & 12.46 & 11.75 & 13.03 \\
Azure Speech CLI 1.37.9 & - & \underline{11.86} & \textbf{10.21}& 11.88 & \underline{11.78} \\
\hline
\textbf{Ours} \\
Zipformer (from scratch)  & 68 & 19.7 & 27.02 & 23.18 & 20.54 \\
VietASR Iteration 1 & 68 & 9.60 & 14.75 & 12.74 & 10.28 \\
VietASR Iteration 2 & 68 & 8.01 & 12.21 & 11.40 & 8.68 \\
VietASR Iteration 3 & 68 & $\textbf{7.68}$ & $\underline{11.46}$ & \underline{10.96} & \textbf{8.31} \\
\bottomrule
\end{tabular}}
\label{tab:main}
\vspace{-1em}
\end{table}

\subsection{Analysis of the Impact of Supervised Codebook}
\label{sec:compare_with_hubert}
To demonstrate the impact of introducing the supervised codebook on pre-training, we compare the performance of the VietASR codebook with the HuBERT codebook, which uses k-means clustering on the spectrum for the first iteration and improves the codebook quality using only unlabeled data.

We re-implement the HuBERT codebook with the following configuration. In HuBERT iteration 1, we apply k-means clustering to the 80-channel Fbank features. In iteration 2, k-means is applied to the hidden states of the third stack of layers of the first iteration's checkpoint, which has the best performance when used as the input of ASR model in our experiment. 


We test the two codebooks in a two-iteration pre-training process. For iteration 1, we run pre-training for 9 epochs, and 18 epochs for iteration 2. The cluster size for k-means is 500.

The Table~\ref{tab:compare} shows the WER of different iterations. For iteration 1, VietASR can provide an average WER reduction of 10.8\% when compared to HuBERT. For iteration 2, VietASR can reduce the average WER by 8.0\% when compared with HuBERT. With only a small additional cost for training the label extractor, VietASR achieves a significant performance improvement over HuBERT.

\begin{table}[th]
\centering
\caption{Comparison between HuBERT codebook and VietASR codebook. The average WER (Avg) is weighted by the word count of each test set.}

\renewcommand{\arraystretch}{1.1}
\resizebox{\linewidth}{!}{
\renewcommand\tabcolsep{5.0pt}
\begin{tabular}{lcccc>{\columncolor[HTML]{F4F4F4}}c}
\toprule
\textbf{System} & \textbf{\makecell{Pre-train \\ Epochs}} & \textbf{\makecell{Giga- \\ Speech 2}}  & \textbf{\makecell{Common \\ Voice}} & \textbf{FLEURS}  & \textbf{Avg} \\
\midrule
\textbf{Iteration 1}  &  &  &  \\
HuBERT Codebook  & 9 & 10.80 & 16.77 & 13.97 & 11.53\\
VietASR  Codebook & 9 &  9.60 & 14.75 & 12.74 & 10.28 \\
\textbf{Iteration 2}  &  &  &  \\ 
HuBERT Codebook & 18 & 8.77 & 13.15 & 12.07 & 9.44 \\
VietASR Codebook & 18 & 8.01 & 12.21 & 11.40 & 8.68 \\
\bottomrule
\end{tabular}}
\label{tab:compare}
\vspace{-1em}
\end{table}

\begin{table}[th]
\centering
\caption{WER for streaming models, the average WER (Avg) is weighted by the word count of each test set.}

\renewcommand{\arraystretch}{1.1}
\renewcommand\tabcolsep{10pt}
\resizebox{\linewidth}{!}{
\begin{tabular}{lccc>{\columncolor[HTML]{F4F4F4}}c}
\toprule
\textbf{Pre-training} & \textbf{\makecell{Giga- \\ Speech 2}} & \textbf{\makecell{Common \\ Voice}} & \textbf{FLEURS} & \textbf{Avg} \\
\midrule
N/A & 21.37 & 30.24 & 27.81 & 22.69 \\
HuBERT Codebook  & 10.70 & 18.38 & 14.78 & 11.64 \\
VietASR Codebook  &  9.26 & 16.49 & 13.04 & 10.13 \\
\bottomrule
\end{tabular}}
\label{tab:streaming}
\vspace{-1em}
\end{table}

\subsection{Performance of Streaming Systems}
We apply VietASR to streaming ASR training. The encoder backbone is replaced with a causal Zipformer. During pre-training and fine-tuning, the chunk size (latency) is randomly chosen among (320ms, 640ms, 1280ms) or set to no chunking. The speech context size is randomly chosen among (1280 ms, 2560 ms, 5120 ms) or full context. The pre-training codebook is the same as the second iteration codebook in Section~\ref{sec:compare_with_hubert}. For decoding, we use a chunk size of 640 ms and a context size of 5120 ms. The result is listed in Table \ref{tab:streaming}. All pre-trained models significantly outperform the baseline, with VietASR achieving the best performance. Specifically, the streaming VietASR model yields an average WER reduction of 13.0\% compared to pretraining with HuBERT codebook, surpassing most open-source and commercial models presented in Table \ref{tab:main}, demonstrating its strong generalization ability in streaming ASR.

%% file: text/limitation.tex

%% file: text/conclusion.tex
\section{Conclusion}
In this paper, we propose VietASR, a training pipeline designed for low-resource ASR, supporting both online and offline scenarios. By adapting the HuBERT architecture to Zipformer and incorporating multi-iteration ASR-biased self-supervised learning, VietASR provides a cost-effective and practical solution for improving ASR performance using limited annotated data and a large-scale unlabeled dataset.
Experiments on Vietnamese demonstrate that VietASR, pre-trained on 70,000-hour unlabeled data and fine-tuned on merely 50-hour labeled data, outperforms the 1542M Whisper Large-v3 and commercial ASR systems on real-world data with only 68M parameters.

While this work is currently limited to Vietnamese, our pipeline is designed for low-resource languages, and we plan to extend it to more languages in the future. All related resource will be open-sourced to facilitate research in low-resource ASR.

%% file: text/acknowledgement.tex
\section{Acknowledgement}
This work was supported by the National Natural Science Foundation of China  (No. 62206171 and No. U23B2018), Shanghai Municipal Science and Technology Major Project under Grant 2021SHZDZX0102, and the Tencent AI Lab Rhino-Bird Focused Research Program.

%% file: main.bbl
\begin{thebibliography}{10}
\providecommand{\url}[1]{#1}
\csname url@samestyle\endcsname
\providecommand{\newblock}{\relax}
\providecommand{\bibinfo}[2]{#2}
\providecommand{\BIBentrySTDinterwordspacing}{\spaceskip=0pt\relax}
\providecommand{\BIBentryALTinterwordstretchfactor}{4}
\providecommand{\BIBentryALTinterwordspacing}{\spaceskip=\fontdimen2\font plus
\BIBentryALTinterwordstretchfactor\fontdimen3\font minus \fontdimen4\font\relax}
\providecommand{\BIBforeignlanguage}[2]{{%
\expandafter\ifx\csname l@#1\endcsname\relax
\typeout{** WARNING: IEEEtran.bst: No hyphenation pattern has been}%
\typeout{** loaded for the language `#1'. Using the pattern for}%
\typeout{** the default language instead.}%
\else
\language=\csname l@#1\endcsname
\fi
#2}}
\providecommand{\BIBdecl}{\relax}
\BIBdecl

\bibitem{CTC}
A.~Graves, S.~Fern{\'a}ndez, F.~Gomez, and J.~Schmidhuber, ``Connectionist temporal classification: labelling unsegmented sequence data with recurrent neural networks,'' in \emph{Proc. ICML}, 2006.

\bibitem{RNN_ASR}
A.~Graves, A.~Mohamed, and G.~Hinton, ``Speech recognition with deep recurrent neural networks,'' in \emph{Proc. ICASSP}, 2013.

\bibitem{li2022recent}
J.~Li \emph{et~al.}, ``Recent advances in end-to-end automatic speech recognition,'' \emph{APSIPA Transactions on Signal and Information Processing}, 2022.

\bibitem{whisper}
A.~Radford, J.~W. Kim, T.~Xu \emph{et~al.}, ``Robust speech recognition via large-scale weak supervision,'' in \emph{Proc. ICML}, 2023.

\bibitem{canary}
K.~C. Puvvada, P.~{\.Z}elasko, H.~Huang \emph{et~al.}, ``Less is more: Accurate speech recognition \& translation without web-scale data,'' \emph{arXiv preprint arXiv:2406.19674}, 2024.

\bibitem{universial-1}
F.~M. Ramirez, L.~Chkhetiani, A.~Ehrenberg \emph{et~al.}, ``Anatomy of industrial scale multilingual {ASR},'' \emph{arXiv preprint arXiv:2404.09841}, 2024.

\bibitem{xls-r}
A.~Babu, C.~Wang, A.~Tjandra \emph{et~al.}, ``{XLS-R}: Self-supervised cross-lingual speech representation learning at scale,'' in \emph{Proc. INTERSPEECH}, 2022.

\bibitem{USM}
Y.~Zhang, W.~Han, J.~Qin \emph{et~al.}, ``{Google USM}: Scaling automatic speech recognition beyond 100 languages,'' \emph{arXiv preprint arXiv:2303.01037}, 2023.

\bibitem{MMS}
V.~Pratap, A.~Tjandra, B.~Shi \emph{et~al.}, ``Scaling speech technology to 1,000+ languages,'' \emph{Journal of Machine Learning Research}, 2024.

\bibitem{yang2024gigaspeech2}
Y.~Yang, Z.~Song, J.~Zhuo \emph{et~al.}, ``{GigaSpeech 2}: An evolving, large-scale and multi-domain {ASR} corpus for low-resource languages with automated crawling, transcription and refinement,'' \emph{arXiv preprint arXiv:2406.11546}, 2024.

\bibitem{Zipformer}
Z.~Yao, L.~Guo, X.~Yang \emph{et~al.}, ``{Zipformer}: A faster and better encoder for automatic speech recognition,'' in \emph{Proc. ICLR}, 2023.

\bibitem{Hubert}
W.-N. Hsu, B.~Bolte, Y.-H.~H. Tsai \emph{et~al.}, ``{HuBERT}: Self-supervised speech representation learning by masked prediction of hidden units,'' \emph{IEEE/ACM Transactions on Audio, Speech, and Language Processing}, 2021.

\bibitem{wav2vec2}
A.~Baevski, Y.~Zhou, A.~Mohamed, and M.~Auli, ``{wav2vec 2.0}: A framework for self-supervised learning of speech representations,'' in \emph{Proc. NeurIPS}, 2020.

\bibitem{PBERT}
C.~Wang, Y.~Wang, Y.~Wu \emph{et~al.}, ``Supervision-guided codebooks for masked prediction in speech pre-training,'' in \emph{Proc. INTERSPEECH}, 2022.

\bibitem{polybert}
Z.~Ma, Z.~Zheng, G.~Yang \emph{et~al.}, ``Pushing the limits of unsupervised unit discovery for {SSL} speech representation,'' in \emph{Proc. INTERSPEECH}, 2023.

\bibitem{HuBERT-AP}
S.~Ren, S.~Liu, Y.~Wu, L.~Zhou, and F.~Wei, ``Speech pre-training with acoustic piece,'' in \emph{Proc. INTERSPEECH}, 2022.

\bibitem{hubert_academic}
W.~Chen, X.~Chang, Y.~Peng \emph{et~al.}, ``Reducing barriers to self-supervised learning: {HuBERT} pre-training with academic compute,'' in \emph{Proc. INTERSPEECH}, 2023.

\bibitem{ASRBERT}
H.~Y. Kim, B.-Y. Kim, S.~W. Yoo \emph{et~al.}, ``{ASBERT}: {ASR}-specific self-supervised learning with self-training,'' in \emph{Proc. SLT}, 2023.

\bibitem{bias_asr}
F.~L. Kreyssig, Y.~Shi, J.~Guo \emph{et~al.}, ``Biased self-supervised learning for {ASR},'' in \emph{Proc. INTERSPEECH}, 2023.

\bibitem{ctcbert}
R.~Fan, Y.~Wang, Y.~Gaur, and J.~Li, ``{CTCBERT}: Advancing hidden-unit {BERT} with {CTC} objectives,'' in \emph{Proc. ICASSP}, 2023.

\bibitem{MelHubert}
T.-Q. Lin, H.-y. Lee, and H.~Tang, ``{MelHuBERT}: A simplified hubert on mel spectrograms,'' in \emph{Proc. ASRU}, 2023.

\bibitem{FastHubert}
G.~Yang, Z.~Ma, Z.~Zheng \emph{et~al.}, ``{Fast-HuBERT}: an efficient training framework for self-supervised speech representation learning,'' in \emph{Proc. ASRU}, 2023.

\bibitem{yang2024k2ssl}
Y.~Yang, J.~Zhuo, Z.~Jin \emph{et~al.}, ``{k2SSL}: A faster and better framework for self-supervised speech representation learning,'' \emph{arXiv preprint arXiv:2411.17100}, 2024.

\bibitem{transformer}
A.~Waswani, N.~Shazeer, N.~Parmar \emph{et~al.}, ``Attention is all you need,'' in \emph{Proc. NeurIPS}, 2017.

\bibitem{conformer}
A.~Gulati, J.~Qin, C.-C. Chiu \emph{et~al.}, ``{Conformer}: Convolution-augmented transformer for speech recognition,'' in \emph{Proc. INTERSPEECH}, 2020.

\bibitem{stateless_predictor}
M.~Ghodsi, X.~Liu, J.~Apfel \emph{et~al.}, ``{RNN-T}ransducer with stateless prediction network,'' in \emph{Proc. ICASSP}, 2020.

\bibitem{transducer}
A.~Graves, ``Sequence transduction with recurrent neural networks,'' \emph{arXiv preprint arXiv:1211.3711}, 2012.

\bibitem{Pruned-RNN-T}
F.~Kuang, L.~Guo, W.~Kang \emph{et~al.}, ``Pruned {RNN-T} for fast, memory-eﬀicient {ASR} training,'' in \emph{Proc. INTERSPEECH}, 2022.

\bibitem{gao2023funasr}
Z.~Gao, Z.~Li, J.~Wang \emph{et~al.}, ``{FunASR}: A fundamental {End-to-End} speech recognition toolkit,'' in \emph{Proc. INTERSPEECH}, 2023.

\bibitem{ardila-etal-2020-common}
R.~Ardila, M.~Branson, K.~Davis \emph{et~al.}, ``{Common Voice}: A massively-multilingual speech corpus,'' in \emph{Proc. ACL}, 2020.

\bibitem{conneau2023fleurs}
A.~Conneau, M.~Ma, S.~Khanuja \emph{et~al.}, ``{FLEURS}: Few-shot learning evaluation of universal representations of speech,'' in \emph{Proc. SLT}, 2023.

\bibitem{BPE}
R.~Sennrich, B.~Haddow, and A.~Birch, ``Neural machine translation of rare words with subword units,'' in \emph{Proc. ACL}, 2016.

\bibitem{modified_beam_search}
W.~Kang, L.~Guo, F.~Kuang \emph{et~al.}, ``Fast and parallel decoding for transducer,'' in \emph{Proc. ICASSP}, 2023.

\end{thebibliography}
